\documentclass[preprint,prd,aps,amsmath,nofootinbib,superscriptaddress,tightenlines]{revtex4}

\usepackage[dvips,bookmarks=false]{hyperref}
\usepackage{graphicx}
\usepackage{bm}
\usepackage{xspace}
\usepackage{slashed}

  \newcount\hour \newcount\minute
  \hour=\time \divide \hour by 60
  \minute=\time
  \newcommand{\mydate}{\ \today \ - \number\hour :\ifnum \minute<10 0\fi 
\number\minute}

\newcommand{\usn}{us/n}
\newcommand{\usbn}{us/\bn}
\newcommand{\usnp}{us/n'}

\def\OMIT#1{}

\newcommand{\nslash}{\slashed{n}}
\newcommand{\bnslash}{\slashed{\bar n}}
\newcommand{\hc}{hc}

\newcommand{\nn}{\nonumber} 

\newcommand{\bn}{{\bar n}}
\newcommand{\bea}{\begin{eqnarray}}
\newcommand{\eea}{\end{eqnarray}}

\newcommand{\cP}{{\cal P}}

\newcommand{\mcdot}{\!\cdot\!}

\newcommand{\SCETa}{\ensuremath{{\rm SCET}_{\rm I}}\xspace}
\newcommand{\SCETb}{\ensuremath{{\rm SCET}_{\rm II}}\xspace}

\begin{document}
\setlength\baselineskip{17pt}


\preprint{ \vbox{ \hbox{MIT-CTP 3790}  \hbox{CMU-HEP-06-09}  
 \hbox{hep-ph/0611356} } 
}

\title{\boldmath 
Three-Parton Contributions to $B\to M_1 M_2$ Annihilation at Leading Order}

\vspace*{1cm}

\author{Christian M.\ Arnesen} 
\affiliation{Center for Theoretical Physics, Laboratory
for Nuclear Science, Massachusetts \\
Institute of Technology, Cambridge, MA 02139}

\author{Ira Z.\ Rothstein}
\affiliation{Department of Physics, Carnegie Mellon University,
  Pittsburgh, PA 15213\vspace*{0.5cm}}

\author{Iain W.\ Stewart\vspace{0.4cm}}
\affiliation{Center for Theoretical Physics, Laboratory
for Nuclear Science, Massachusetts \\
Institute of Technology, Cambridge, MA 02139}

\begin{abstract}\vspace*{0.2cm}
  
  We compute annihilation amplitudes for charmless $B$ decays that are
  proportional to the three-parton twist-3 light meson distribution
  amplitude $\phi_{3M}(x_1,x_2)$ with an active gluon. Due to an
  enhancement from a quark propagator at the scale $p^2\sim
  m_b\Lambda_{\text{QCD}}$ these terms occur at the same parametric order in
  $\alpha_s(m_b)$ and $1/m_b$ as the known leading order annihilation
  involving $f_B$ and twist-2 meson distributions. With our
  calculation the leading order annihilation amplitude is now
  complete.  At lowest order in $\alpha_s$ the new amplitudes are real and only $O_{5-8}$
  contribute. Using simple models we find that the three-parton and
  two-parton terms are of comparable size.

\end{abstract}

\maketitle

The nonleptonic charmless decay channels \(B \to M_1 M_2\) provide a wealth of
information about the standard model, including the study of \(CP\) violation
and the strong interactions.  Since many amplitudes for these decays are loop
dominated, it is possible for new physics to give a significant contribution.
However, except for the simplest observables, testing for new physics requires
an understanding of the standard model background.  Predicting standard model
decay rates and $CP$ asymmetries with quantum chromodynamics (QCD) is difficult,
but the task is simplified by the use of the soft-collinear effective theory
(SCET)~\cite{Bauer:2000ew,Bauer:2000yr,Bauer:2001ct,Bauer:2001yt} and
factorization
theorems~\cite{Beneke:2000ry,Keum:2000ph,Chay:2003zp,Bauer:2004tj}.

An interesting experimental observable is the relative ``strong''
phase between standard model amplitudes multiplying the CKM factors
$V_{ub}V_{uf}^*$ and $V_{cb}V_{cf}^*$ ($f=d,s$), since these phases
are measured to be large in the $B\to \pi\pi$ and $B\to K\pi$
channels~\cite{CKMfitter}. There are two competing standard model
explanations for these phases, sizeable charm penguin
loops~\cite{Ciuchini:1997hb,Colangelo:1989gi,Bauer:2004tj,Williamson:2006hb}
or sizeable annihilation
amplitudes~\cite{Keum:2000ph,Lu:2000em,Beneke:2001ev,Kagan:2004uw,Khodjamirian:2005wn}
in which the initial state ``spectator'' quark is Wick-contracted with
a quark field in the effective Hamiltonian.  In this paper we report
on a new contribution to the leading annihilation amplitudes.

Our notation follows that of Ref.~\cite{Bauer:2004tj} where factorization
theorems for the leading order $B\to M_1M_2$ amplitudes were derived with an
expansion in $\Lambda/Q$ where $\Lambda$ is a hadronic scale and $Q\sim m_b \sim
m_c \sim E_M$.  We restrict our discussion to non-isosinglet mesons ($M_i=$
$\pi$, $K$, $\rho$, $\ldots$) which can not be produced solely by gluons, 
for which the annihilation amplitudes are power suppressed by \(\sim\Lambda/Q\).
Recently these power corrections were classified according to their perturbative
order and source for strong phases~\cite{Arnesen:2006vb}.  The annihilation
ampltude is
\begin{align}\label{annexpansion}
  A^{ann}&=A^{(1ann)}_{\rm hard}+A^{(1ann)}_{\rm hard-col}+A^{(2ann)}_{\rm
    hard}+\ldots \,,
\end{align}
where the superscript denotes the order in $\Lambda/Q$. A subscript ``hard''
denotes annihilation amplitudes generated by propagators offshell by $p^2\sim
m_b^2$.  At lowest order in $\alpha_s$ these include the standard leading order
amplitudes $A^{(1ann)}_{\rm hard}\sim \big[ \alpha_s(m_b) f_B f_{M_1} f_{M_2}
\phi_{M_1} \phi_{M_2}\big]$ as well as the chirally enhanced amplitudes
$A^{(2ann)}_{\rm hard}\sim \big[ \alpha_s(m_b) f_B f_{M_1} f_{M_2} \phi_{M_1}
\phi_{M_2}^{pp}\mu_{M_2}/m_b\big]$. These have been studied in earlier
analyses~\cite{Keum:2000ph,Lu:2000em,Beneke:2001ev,Kagan:2004uw}, and at this
order, factorization in rapidity~\cite{Manohar:2006nz} reveals that they are
real~\cite{Arnesen:2006vb}. Here the $f$'s are decay constants, and the chiral
enhancement factors are $\mu_\pi = m_\pi^2/(m_u + m_d)$ and $\mu_K = m_K^2/(m_u
+ m_s)$. $\phi_{M_i}$ are twist-2 distribution functions, and $\phi_{M}^{pp}$ is
a two-parton twist-3 distribution. In Eq.~(\ref{annexpansion}) the amplitudes
$A_{\rm hard-col}^{(1ann)}$ have hard-collinear propagators, which are offshell
by an amount $\mu_i^2\sim m_b\Lambda$. A non-perturbative phase is first
generated by soft exchange between the two mesons as an order
$\alpha_s^2(\mu_i)/\pi$ suppressed term in $A_{\rm hard-col}^{(1ann)}$. The
ellipsis in Eq.~(\ref{annexpansion}) denotes the fact that the full set of
$A^{(2ann)}$ amplitudes have not yet been classified.

In this paper we compute the leading term in the perturbative expansion of $A_{\rm
  hard-col}^{(1ann)}$, which has the form
\begin{align} \label{Anew}
A_{\rm hard-col}^{(1ann)}\sim
  \alpha_s(m_b) \frac{H(x_1,y_1,y_2)}{k}\otimes  f_B \phi_B^+(k) \:  f_{M_1}
  \phi_{M_1}(x_1)  \:
  f_{3M_2} \phi_{3M_2}(y_1,y_2) \,.
\end{align}
Here $H$ is a calculable hard-scattering kernel, $\phi_{3M}$ is a
three-parton twist-3 distribution, and $f_{3M}$ is the corresponding
decay constant.  The amplitude in Eq.~(\ref{Anew}) occurs at the same
order in $1/m_b$ and $\alpha_s(m_b)$ as $A_{\rm hard}^{(1ann)}$ and
should be included for a complete leading order annihilation
amplitude. Unlike $A_{\rm hard}^{(1ann)}$ its convolution integrals
converge without using rapidity factorization. Furthermore, the LO annihilation involves \(B\)-meson information beyond \(f_B\), thus demonstrating that
annihilation is more complicated than the short distance picture leading to a scaling $\sim
f_B/m_b$
that is often used in parametric estimates~\cite{Blok:1997yj}.

In QCD at the scale \(m_b\), flavor changes are mediated by the weak effective Hamiltonian. For $B\to M_1M_2$ with $\Delta S=0$,
\begin{equation} \label{Hw}
H_W = \frac{G_F}{\sqrt{2}} \sum_{p=u,c} V_{pb} V^*_{pd}\,
  \Big( C_1 O_1^p + C_2 O_2^p 
  + \!\sum_{i=3}^{10,7\gamma,8g}\! C_i O_i \Big).  
\end{equation} 
Most of these operators have spin $(V\!-\!A)\otimes(V\!-\!A)$, such as $O_1^u = (\bar u
b)_{V\!-\!A}\,(\bar d u)_{V\!-\!A}$. We will prove below that all such operators give vanishing contribution to Eq.~(\ref{Anew}), so that only
\begin{align}\label{fullops}
 O_{5}  &= {\textstyle \sum}_{q'} (\bar d b)_{V\!-\!A}\,
  (\bar q' q')_{V\! + \!A} \,, 
 & O_6 &= {\textstyle \sum}_{q'} (\bar d_{\beta} b_{\alpha})_{V\!-\!A}\,
  (\bar q'_{\alpha} q'_{\beta})_{V\! + \!A} \,, \nn \\
 O_{7}  &= {\textstyle \sum}_{q'} \frac{3e_{q'}}{2}\, (\bar d b)_{V\!-\!A}\,
  (\bar q' q')_{V\! + \!A}\,,
 & O_8 & = {\textstyle \sum}_{q'} \frac{3e_{q'}}{2}\,
  (\bar d_{\beta} b_{\alpha})_{V\!-\!A}\,
  (\bar q'_{\alpha} q'_{\beta})_{V\! + \!A} \,, 
\end{align}
are relevant for our analysis. Here $\alpha$ and $\beta$ are color indices, $e_{q'}$ are electric charges,  and the sum
is over flavors $q'=u,d,s,c,b$. Results for $\Delta
S=1$ transitions are obtained by replacing $d\to s$ in Eqs.~(\ref{Hw}) and
(\ref{fullops}), and likewise in the equations below.  The coefficients in
Eq.~(\ref{Hw}) are known at next-to-leading-log order~\cite{Buchalla:1995vs}. (We have
$O^p_{1}\leftrightarrow O^p_{2}$ relative to~\cite{Buchalla:1995vs}). In the NDR
scheme with $m_b=4.8\,{\rm GeV}$, the coefficients are
$C_{5-8}(m_b) = \big\{
  0.010,
 -0.040 , 
  4.9 \!\times\! 10^{-4} ,
  4.6 \!\times\! 10^{-4} 
  \big\}, $
They are considerably smaller than $C_1(m_b)=1.08$ and $C_2(m_b)=-0.18$, but can give important contributions in penguin observables because $C_{1,2}$ only contribute through loops~\cite{Beneke:2001ev}.

To separate the mass scales below $m_b$ we match $H_W$ onto operators
in SCET.  The amplitude for \(B \to M_1 M_2\) is most easily calculated in the
\(B\) rest frame where soft fields with typical momenta \(\sim\Lambda\) interpolate for the initial state \(B\).
The final state hadrons $M_1$ and \(M_2\) are back-to-back energetic charmless
pseudoscalar or vector mesons.  Collinear fields in the light-like direction
\(n\) interpolate for one light meson, and collinear fields in the direction
\(\bar n\) interpolate for the other. These fields have typical momenta
\((n\cdot p,\bar n \cdot p, p_\bot)\sim Q(\eta^2,1,\eta) \) and
\(Q(1,\eta^2,\eta)\), respectively, in terms of the power counting parameter
$\eta\sim \Lambda/Q$. The vectors \(n\) and \(\bar n\) satisfy \(n\cdot \bar
n=2\), and we work in a frame where the $B$-meson four-velocity \(v\) has
\(n\cdot v=\bn \cdot v=1\).  To calculate the amplitude in Eq.~(\ref{Anew}) we
first match QCD onto operators in \SCETa where the hard-collinear modes with
$p^2\sim m_b\Lambda$ are still propagating and then match these onto operators
in \SCETb which has only non-perturbative modes with $p^2\sim \Lambda^2$~\cite{Bauer:2002aj}.

Before presenting the details of the calculation of Eq.~(\ref{Anew})
we complete our review of Ref.~\cite{Arnesen:2006vb} with a discussion
of the required ingredients and an overview of the matching
procedure. Let $Q^{(0)}$ denote the leading \SCETa weak operators.
Annihilation contributions require either (a) an $n$-quark
\(\bn\)-antiquark pair produced by hard interactions giving a
$Q^{(k\ge 2)}$ six-quark operator, or (b) a time-ordered product of
two mixed Lagrangians ${\cal L}_{\xi_n q}^{(k)}{\cal L}_{\xi_{\bn}
q}^{(k)}$ to produce the $n$-$\bn$ pair by soft exchange. Annihilation
also requires a mechanism for connecting the ``spectator'' to the weak
operator, either by (i) having a soft field directly in the weak
operator, or (ii) having a time-ordered product with an ${\cal L}_{\xi
q}^{(k)}$. Case (a,i) gives operators $Q^{(4)}$ which contribute only
to $A_{\rm hard}^{(1ann)}$. Case (b,i) vanishes at this order.  Case
(b,ii) involves three hard-collinear gluons and is $\sim
\alpha_s^2(\mu_i)$.  This leaves case (a,ii). Here $Q^{(2)} {\cal
L}_{\xi q}^{(1)}$ can contribute at ${\cal O}(\alpha_s(m_b))$ if the
gluon from ${\cal L}_{\xi q}^{(1)}$ is uncontracted.  Since the
uncontracted gluon costs an extra power when matched onto \SCETb, it
is only this class of operators that contributes with an external
gluon, not $Q^{(3)}{\cal L}_{\xi q}^{(1)}$, $Q^{(2)} {\cal L}_{\xi
q}^{(2)}$, etc.
\begin{figure}
  \centerline{ 
   \includegraphics[width=7cm]{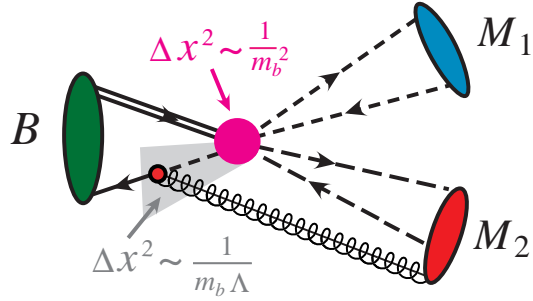}
  }
  \vskip-0.2cm
\caption{Picture for  $Q_i^{(2)}{\cal L}^{(1)}_{\xi q}$,
  which generates an annihilation amplitude that is sensitive to the
  intermediate scale $p^2 \sim m_b\Lambda$. The filled circle at the
  center represents a \SCETa six-quark operator \(Q^{(2)}\) arising
  from the full-theory diagrams in Fig.~(\ref{fig:Ann}) at the hard
  scale \(p^2\sim m_b^2\).  \label{fig:AnnFACT}}
\end{figure}
Thus the calculation of \(A^{(1ann)}_{hard-col}\) involves finding
\SCETa operators of the form
\begin{equation} \label{Q2early}
  Q_{id}^{(2)}\propto \big[ \bar q'_{n',\omega_5}\Theta_{us} b_v \big]
    \big[ \bar d_{\bar n,\omega_2} \Theta _{\bar n} q_{\bar n,\omega_3} \big]
    \big[ \bar q_{ n,\omega_1} \Theta _{n} q'_{ n,\omega_4} \big]\,,
\end{equation}
where $\Theta_{us}\otimes \Theta_\bn \otimes \Theta_n$ are color and
spin structures, $q$ and $q'$ are flavors, and the collinear direction
\(n'=n\) or
\(\bar n\).  The fermion fields are gauge invariant with large label momenta specified by the subscripts \(\omega\), for example $q_{n,\omega_1}=\delta(\omega_1-\bn\mcdot\cP)W^\dagger_n
\xi_n^{(q)}$ where $W_n$ is a Wilson line.  At tree level these operators arise
from the full-theory diagrams in Fig.  \ref{fig:Ann} with three light
\(n'\)-collinear quarks and two collinear in the other direction,
\(\bn'\). They have Wilson coefficients of \(\mathcal
O(\alpha_s(m_b))\). We identify \(n\) as the collinear direction of
the pair-produced quark of flavor $q$ and sum over all \(n\) in the
\SCETa weak Hamiltonian.  We will see shortly that the flavor
structure is as in Eq.~(\ref{Q2early}), and that the matching requires
$T^A$ color structures for two of the $\Theta$'s.

\begin{figure}[t]
\includegraphics[width=16cm]{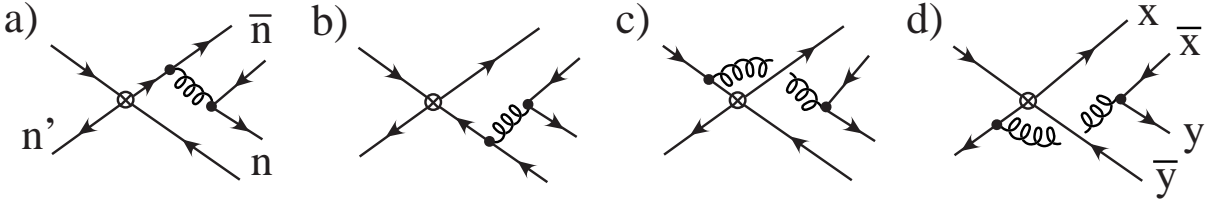}
\vskip-0.4cm
\caption{Tree-level annihilation graphs for $B\to M_1 M_2$ decays. The gluon and the fermion propagator connecting it to the weak vertex are both offshell by \(p^2\sim m_b\). Matching on to \SCETa, these graphs give rise to the six-quark operators \(Q^{(2)}\), the filled circle at the center of Fig.~\ref{fig:AnnFACT}.
  \label{fig:Ann}}
\end{figure}

To finalize our description of the calculation we consider matching the
time-ordered product \(T[Q^{(2)} \mathcal L_{\xi q}^{(1)}]\) onto
\SCETb with diagrams as shown in Figure \ref{fig:AnnFACT}. \(Q^{(2)}\) has an excess of \(n'\)-collinear fermions since only two
are needed to interpolate for a collinear meson. The
subleading Lagrangian~\cite{Beneke:2002ni}
$\mathcal L^{(1)} _{\xi q}
   = \bar q'_{us}  ig \slashed {\mathcal B} _{n'}^\perp q'_{n'}$ 
removes an \(n'\)-collinear fermion and provides the soft field that
interpolates for the light anti-quark in the \(B\) meson. Here 
$ig\,{\cal B}^{\perp\,\mu}_{n',\omega} = \big[ 1/(\bn^\prime \mcdot \cP)\, W^\dagger_{n'} 
[ i\bn\cdot D_{n'} , i D^\mu_{n',\perp} ] W_{n'} \delta(\omega-\bn'\mcdot \cP^\dagger) \big]$, and the form of the \SCETb operators is
\begin{equation} \label{O1Tearly}
  O_{id}^{(1T)} \propto \frac{1}{n'\!\cdot\! k} \:
    \big[ \bar q'_{s,n'\!\cdot k} \Gamma_s b_v \big]
    \big[ \bar d_{\bn} \Gamma_{\bn} q_{\bn} \big]
    \big[ \bar q_n \Gamma_n q'_n \big]  ig\mathcal B_{n'}^{\perp\beta} \,,
\end{equation} 
with $\Gamma_s\otimes \Gamma_{\bn}\otimes\Gamma_n$ containing spin and color
structures. The collinear gluon field strength \(ig\mathcal B_{n'}^{\perp}\sim\eta\),
interpolates for gluons in a final state meson, so there is no perturbative
suppression from the factor of $g$.
 At tree level, integrating out the hard-collinear quark propagator
in Fig.~\ref{fig:AnnFACT} induces an inverse factor $1/(n'\cdot k)$ of the soft
momentum which will be convoluted with the $B$-distribution, $\phi_B^+(n'\cdot
k)$. In Eq.~(\ref{O1Tearly}) this compensates the \(\eta\) suppression from
\(ig\mathcal B_{n'}^{\perp}\) to make $O^{(1T)}$ the same order as the six-quark
operators for the hard annihilation, which is ${\cal O}(\eta^7)$. We
have checked that operators with more
\(ig\mathcal B_{n'}^{\perp}\)'s or with soft gluon field strengths do not occur
at this order in $1/m_b$ and $\alpha_s(m_b)$. 

Note that \SCETb time-ordered products (T-products) do not contribute
at ${\cal O}(\eta^7)$. To see this, recall that our process has a soft
initial state and $n$ and $\bn$-collinear final states. An example of
an \SCETb Lagrangian that connects these sectors~\cite{Hill:2002vw}
has two-collinear quarks and two-soft quarks~\cite{Mantry:2003uz},
$\bar q_s q_s \bar \xi_n \xi_n\sim \eta$.
  In these operators the two $n$-collinear
particles conserve the large $p^-\sim \eta^0$ momenta, and the two
soft particles conserve the $p^+\sim \eta$ momenta. Thus this
operator, as well as analogous operators with gluons, only support
scattering, $ns \to ns$, and not annhilation such as $nn\to ss$ or
$ss\to nn$.  Another example is ${\cal L}_{\rm II}\sim \bar \xi_n A_n A_\bn
\xi_\bn\sim\eta^2$, where analogous statements hold for $n$ and
$\bn$. Weak operators, like $O_{id}^{(1T)}$, that have the same
$n$-$\bn$-$s$ structure as the initial and final states are already
${\cal O}(\eta^7)$, so T-products with them are power suppressed.
 The above considerations rule out the
majority of T-products. An example of an annihilation T-product in
\SCETb that survives these criteria is ${\cal L}_{\rm II}$, with a
weak operator with fields $(\bar q_s h_v \bar\xi_n \xi_\bn)\sim
\eta^5$.  These T-products involve at least one loop momentum
$\ell^\mu$ where, due to the double multipole expansion, $\ell^\pm$
must be smaller than the conserved $p^-$ and $p^+$, see
Eq.(25) of Ref.~\cite{Stewart:2003gt}.  As a contour integral in $\ell^+$ or
$\ell^-$ we have $\ge 2$ poles that are all on the same side of the
axis, and therefore the loop gives zero. At ${\cal O}(\eta^7)$ this is
sufficient to rule out possible annihilation T-products, including
those with more than one \SCETb Lagrangian.  Note that in
Ref.~\cite{Mantry:2003uz} a T-product contribution was identified 
for $\bar B^0\to D^0\pi^0$, however in that scattering process the 
integral did not satisfy the same pole criteria as we find here.

\vskip0.2cm
\noindent {\bf Constructing the Operator Bases}

 Next we construct a full basis for the operators \(Q^{(2)}\) and \(O^{(1T)}\) in the
\SCETa and \SCETb weak effective Hamiltonians, respectively. (The
matching calculations beginning on page 7 can be understood without
the details of this somewhat technical construction, the results of
which are Eqs.~(\ref{Q2})~and~(\ref{O1T}).) General symmetry arguments
allow us to reduce the operator bases to the small subset relevant to
our calculation of
\(A^{(1ann)}_{\rm hard-col}\), and for this reason it is convenient to
construct the bases for \SCETa and \SCETb simultaneously. First consider spin in
\SCETa. For light fermion fields of definite handedness, a
complete basis of Dirac structures for the individual bilinears in
Eq.~(\ref{Q2early}) is
\begin{equation} \label{basis}
  \Theta_{\usnp} =\{1,\gamma_\bot^\alpha\}\,, \qquad
  \Theta_\bn = \{ \nslash, \nslash\gamma_\perp^\mu \} \,, \qquad
  \Theta_n = \{ \bnslash, \bnslash\gamma_\perp^\mu \} \,. 
\end{equation}
Using these bases, we must construct a complete set of \(Q^{(2)}\)
spin structures with chiralities inherited in perturbative matching
from the full-theory fields in
\(O_{1-10}\) and the produced \(q\bar q\) pair. To make a Lorentz scalar, the spin structure must have zero \(\gamma_\bot\)'s or two \(\gamma_\bot\)'s contracted with \(g_\bot^{\alpha\beta}=g^{\alpha\beta}-n^\alpha\bn^\beta/2-n^\beta\bn^\alpha/2\). Note that contracting with \(\epsilon_\bot^{\alpha\beta}=\bn^\rho n^\sigma\epsilon^{\alpha\beta\rho\sigma}/2\) does not yield an independent operator since for
example $i\epsilon_\perp^{\mu\nu} \bar \xi_n^L \bnslash \gamma^\perp_\nu \xi_n^R
= \bar \xi_n^L \bnslash \gamma_\perp^\mu\gamma_5 \xi_n^R = \bar \xi_n^L \bnslash
\gamma_\perp^\mu \xi_n^R$. For
$O_{1-4,9,10}$ the only allowed chiral structure is $(LH)(LL)(LL)$
where $L$ and $R$ refer to the handedness for the light quarks in the
bilinears in the order shown in Eq.~(\ref{Q2early}).  We cannot assign
a handedness to the heavy quark denoted here by $H$. This chiral
structure is realized as the spin structures
\begin{align}\label{Thetanogps}
\Theta_{\usn}\otimes\Theta_\bn\otimes\Theta_n
&=1\otimes\nslash\otimes\bnslash\,,
&
\Theta_{\usbn}\otimes\Theta_\bn\otimes\Theta_n
&=1\otimes\nslash\otimes\bnslash\,.
\end{align}
We have ruled out the chirality \((LH)(LR)(RL)\) corresponding to a
spin structure \(1\otimes \nslash \gamma_\bot^\alpha\otimes
\bnslash\gamma^\bot_\alpha\) by using
$P_R\bnslash'\gamma_\perp^\alpha\otimes P_L\nslash'\gamma^\perp_\alpha
= 0$. This equation encodes the helicity flip argument of
Ref.~\cite{Kagan:2004uw}.  Similarly, for $O_{5-8}$ the chirality
$(LH)(RR)(RR)$ is also realized as the spin structures
Eq.~(\ref{Thetanogps}), whereas \((LH)(RL)(LR)\) is not allowed since
\(P_L\bnslash\gamma_\perp^\alpha\otimes P_R\nslash\gamma^\perp_\alpha
=0\). We will show momentarily, however, that using \SCETb the terms
in Eq.~(\ref{Thetanogps}) are not needed to compute
\(A^{(1ann)}_{hard-col}\). For $O_{5-8}$ we can also have
\begin{align}\label{Thetatwogps}
\Theta_{\usn}\otimes\Theta_\bn\otimes\Theta_n
&=\gamma_\perp^\alpha \otimes \nslash \otimes \bnslash \gamma^\perp_\alpha\,,
&
\Theta_{\usbn}\otimes\Theta_\bn\otimes\Theta_n
&=\gamma_\perp^\alpha \otimes \nslash \gamma^\perp_\alpha \otimes \bnslash\,,
\end{align}
corresponding to chiralities $(RH)(LL)(LR)$ and $(RH)(LR)(RR)$,
respectively, and thus the flavor structure shown in
Eq.~(\ref{Q2early}), namely $(\bar q' b)(\bar d q)(\bar q q')$. The
second structure in Eq.~(\ref{Thetanogps}) is related to the second
structure in Eq.~(\ref{Thetatwogps}) by a Fierz transformation
swapping $\bar d_\bn$ and $\bar q'_\bn$ quarks and we will choose the
latter for our operator basis. The complete set of spin structures in
Eqs.~(\ref{Thetanogps})~and~(\ref{Thetatwogps}) contains neither
\(\Theta_{\usn}\otimes\Theta_{\bn}\otimes\Theta_{n}=\gamma_\bot^\alpha
\otimes \nslash \gamma^\bot_\alpha \otimes \bnslash\) nor
\(\Theta_{\usbn}\otimes\Theta_{\bn}\otimes\Theta_{n}=\gamma_\bot^\alpha
\otimes \nslash \otimes \bnslash\gamma^\bot_\alpha\). These
possibilities are excluded by the projection relation
\(\Theta_{\usnp}\doteq\bnslash'\nslash'\Theta_{\usnp}\slashed{v}/4\)
and the helicity flip equation.

Now consider spin and chirality in \SCETb. The allowed \(O^{(1T)}\)
spin structures must respect the handedness inherited from the \SCETa
fields in the perturbative matching of \(T [Q^{(2)} \mathcal L _{\xi
q}^{(1)}]\). For \(n'=\bn\) in \(Q^{(2)}\), taking either one of
the \(\bn\)-collinear anti-quark fields soft yields an annihilation
operator. For \(n'=n\), however, the field \(\bar q_n\) in the third
bilinear was pair produced and does not contribute to the annihilation
amplitude when made soft by \( \mathcal L _{\xi q}^{(1)}\). So given
the \SCETa spin structures
Eqs.~(\ref{Thetanogps})~and~(\ref{Thetatwogps}) corresponding to
chiralities described in the text, we need to consider \(O^{(1T)}\)
chiralities \((LH)(LL)(LL),\, (LH)(RR)(RR),\, (RH)(LL)(LR),\) and
\((RH)(LR)(RR)\) with bilinears in the order shown in
Eq.~(\ref{O1Tearly}), \(i.e.\ \text{soft}-\bn-n\). With the first
bilinear purely soft, a complete basis of Dirac structures for the
individual bilinears is
\begin{equation} \label{Gammabasis}
  \Gamma_{s} =\{\nslash,\bnslash,\gamma_\bot^\alpha\}\,, \qquad
  \Gamma_\bn = \{ \nslash, \nslash\gamma_\perp^\mu \} \,, \qquad
  \Gamma_n = \{ \bnslash, \bnslash\gamma_\perp^\mu \} \,. 
\end{equation}
A Lorentz scalar \(O ^{(1T)}\) has an odd number of \(\gamma_\bot\)'s since one must be contracted into the \(n-\) or \(\bn-\)collinear field strength \(\mathcal B^\beta_\bot\). For chiralities \((LH)(LL)(LL)\) and \((LH)(RR)(RR)\) the allowed Dirac structure is
\begin{align} \label{gamma3b}
(\Gamma_s \otimes \Gamma_\bn \otimes \Gamma_n) \mathcal
B^\beta_{n',\perp} &= (\gamma^\perp_\beta \otimes \nslash \otimes
\bnslash )\mathcal B^\beta_{n',\perp} \end{align} with \(n'=n\) or
\(\bn\), but the corresponding operators \(O^{(1T)}\) have $\bar q_s
\gamma_\perp^\mu b_v$ and do not contribute for $B$ decays. Since
\((LH)(LL)(LL)\) is the only \(O^{(1T)}\) chirality corresponding to
the \((V\!-\!A)(V\!-\!A)\) operators \(O_{1-4,9,10}\), this proves
that only $O_{5-8}$ can contribute to Eq.~(\ref{Anew}). Furthermore
since all $(LH)$ terms are ruled out, the soft quark can only be $q'$,
and not a $d$-quark.

This leaves the $(RH)(LL)(LR)$ and $(RH)(LR)(RR)$ structures from $O_{5-8}$ with
soft quark flavor $q'$,  for which we have the additional spin structures,
\begin{align} \label{gamma2b}
  n' =n: & & \Gamma_s \otimes \Gamma_\bn \otimes \Gamma_n \mathcal B^\beta_{n,\perp}
   = \big\{ \nslash \otimes \nslash \otimes \bnslash\: 
       \slash\!\!\!\!\mathcal B_{n,\perp} ,\ 
    \nslash \otimes \nslash\gamma^\perp_\beta \otimes \bnslash 
       \mathcal B_{n,\perp}^\beta \big\} 
   \,,\nn\\
  n'=\bn: &  & \Gamma_s \otimes \Gamma_\bn \mathcal B^\beta_{\bar n ,\perp} \otimes
       \Gamma_n 
   = \big\{  
    \bnslash \otimes \nslash\: \slash\!\!\!\!\mathcal B_{\bar n ,\perp}\otimes
      \bnslash ,\ \bnslash \otimes \nslash \mathcal B_{\bar n ,\perp}^\beta
      \otimes \bnslash\gamma^\perp_\beta   \big\} 
   \,,
\end{align}
plus those with \(\nslash\leftrightarrow\bnslash\) in
\(\Gamma_s\). While these eight are all allowed by chirality and
Lorentz invariance, six can be ruled out by considering the spin and
factorization properties of our time-ordered product. The matching
from \SCETa to \SCETb does not affect the spin and color structure of
the $\bn' $-collinear bilinear at this order in the power expansion,
since once a jet direction is chosen the collinear fields in the
opposite direction are decoupled. Here $\bn'$ is the opposite of $n'$. From
Eqs.~(\ref{Thetanogps})~and~(\ref{Thetatwogps}) the allowed
\(\Theta_{\bn'}\) structures have no
\(\gamma_\perp\)'s, and therefore the second structure on each line of
Eq.~(\ref{gamma2b}) does not appear at any order in the perturbative matching.
Also, the allowed structures Eqs.~(\ref{Thetanogps})~and~(\ref{Thetatwogps}) are invariant under
\(\Theta_{us}\to\Theta_{us}\bnslash'/2 \) and only power-suppressed interactions
couple the \(b-\)quark to the \(n'\) sector.  Therefore, \(\Gamma_s\) should not
vanish under \(\Gamma_{s}\to\Gamma_{s}\bnslash'/2 \), and the operators with
\(\nslash\leftrightarrow\bnslash\) mentioned below Eq.~(\ref{gamma2b}) are
ruled out. In perturbation theory this just corresponds to the appearance of an
$\nslash'$ from the $n'$-collinear propagator next to the $b$-quark. This leaves
only the operators with a $ \slash\!\!\!\!\mathcal B_{\perp}$ in
Eq.~(\ref{gamma2b}).

Finally consider color. In \SCETa the operators \(Q^{(2)}\) are color singlets, but
each bilinear on its own could be singlet or octet. A complete set of color
structures includes
\begin{align}\label{color}
\Theta_{\usnp}\otimes\Theta_\bn\otimes\Theta_n
      =\big\{&T^a \otimes 1 \otimes T^a ,
       \ T^a \otimes T^a \otimes 1,\ 
       1 \otimes 1 \otimes 1, 
       \nn\\ 
 & 1 \otimes T^a \otimes T^a,\ 
  T^a\otimes T^b\otimes T^c f^{abc},\ 
  T^a\otimes T^b\otimes T^c d^{abc}
  \big\}\,.
\end{align}
Once again we can reduce this set using the factorization properties of \SCETa.
As argued for spin, an \SCETa operator with color structure \(\Theta_{\bn'}\)
matches onto a \SCETb operator with the same structure \(\Gamma_{\bn'}\) in its
\(\bn'\) bilinear. So \(\Theta_{\bn'}\) cannot be a color octet, and the allowed
structures are
\begin{align}\label{Thetacolor}
\Theta_{\usn}\otimes\Theta_\bn\otimes\Theta_n
&=\{1\otimes1\otimes1,\, T^a \otimes 1 \otimes T^a \}
\nn\\
\Theta_{\usbn}\otimes\Theta_\bn\otimes\Theta_n
&=\{1\otimes1\otimes1,\, T^a \otimes T^a \otimes 1\}\,.
\end{align}
In \SCETb each of the three bilinears
interpolates for a color singlet meson and therefore each bilinear must
seperately be a color singlet,  $\Gamma_s \otimes \Gamma_\bn \otimes \Gamma_n=1\otimes1\otimes1$.

\vskip0.2cm
\noindent {\bf Matching onto \SCETa and \SCETb}

We now present the matching from $H_W$ in Eq.~(\ref{Hw}) onto the \SCETa operators  \(Q^{(2)}\) and then the matching of the \SCETa time-ordered product \(T[Q^{(2)} \mathcal L_{\xi q}^{(1)}]\) onto \SCETb operators $O_{id}^{(1T)}$.   The hadronic matrix elements of $O_{id}^{(1T)}$ will give the factorization formula for \(A^{(1ann)}_{\rm hard-col}\).  From the arguments presented above, the complete basis of \SCETa operators \(Q^{(2)}\) is
\begin{align} \label{Q2}
  Q_{1d}^{(2)} &= \frac{2}{m_b^3}\, {\textstyle \sum}_{q,q'} \,
   \big[ \bar q'_{n,\omega_5}P_L  \gamma^\alpha_\bot T^a {b}_{v} \big]
  \big[ \bar d_{\bn,\omega_2}  \nslash P_L\, q_{\bn,\omega_3} \big]
  \big[ \bar q_{n,\omega_1}  \bnslash\: \gamma_\alpha^\bot T^a\! P_R\,
      q'_{n,\omega_4} \big]
   \,, \nn \\ 
  Q_{2d}^{(2)} &= \frac{2}{m_b^3}\, {\textstyle \sum}_{q,q'}  \,
   \big[ \bar q'_{\bn,\omega_5}P_L \gamma_\bot^\alpha T^a {b}_{v} \big]
  \big[ \bar d_{\bn,\omega_2}  \nslash \gamma^\bot_\alpha T^a\:  P_R\,
      q_{\bn,\omega_3} \big]
  \big[ \bar q_{n,\omega_1}  \bnslash P_R\, q'_{n,\omega_4} \big]
   \,, \nn \\[5pt]
  Q_{3d,4d}^{(2)} &=  Q_{1d,2d}^{(2)}\,\, \frac{3e_{q'}}{2} \,,
\end{align}
with sums over $q,q'=u,d,s$, plus analogous operators $Q_{5d-8d}^{(2)}$ which
have color structure \(1\otimes1\otimes1\).  The electroweak penguin
operators $O_{7,8}$ induce the two operators $Q_{3d,4d}^{(2)}$, which have the
same spin and flavor structures as $O_{1d,2d}^{(2)}$, but with a factor of the quark electric charge $e_{q'}$ included under the summation. 
Combining the pieces in \SCETb, a complete basis for the ${\cal O}(\eta^7)$
operators with one $i g \mathcal B_\perp^\beta$ that contribute to \(B\) decays
is
\begin{align} \label{O1T}
  O_{1d}^{(1T)} &= \frac{1}{m_b^3\, k^+}\, {\textstyle \sum}_{q,q'} \,
   \big[ \bar q'_{s,-k^+}P_L \nslash\, S_n^\dagger {b}_{v} \big]
  \big[ \bar d_{\bn,\omega_2}  \nslash P_L\, q_{\bn,\omega_3} \big]
  \big[ \bar q_{n,\omega_1}  \bnslash\: (ig\slashed{\mathcal B}_\perp)_{n,\omega_5}
   P_R\,   q'_{n,\omega_4} \big]
   \,, \nn \\ 
  O_{2d}^{(1T)} &= \frac{1}{m_b^3\, k^-}\, {\textstyle \sum}_{q,q'}  \,
   \big[ \bar q'_{s,-k^-}P_L  \bnslash\,  S_\bn^\dagger {b}_{v} \big]
  \big[ \bar d_{\bn,\omega_2}  \nslash\: (ig\slashed{\mathcal
    B}_\perp)_{\bn,\omega_5} P_R\,
      q_{\bn,\omega_3} \big]
  \big[ \bar q_{n,\omega_1}  \bnslash P_R\, q'_{n,\omega_4} \big]
   \,, \nn \\[5pt]
  O_{3-4d}^{(1T)} &= O_{1-2d}^{(1T)} \ \frac{3e_{q'}}{2} \,.
\end{align}
Here \(\bar q'_{s,-k^+}=(\bar q'_{s} S_n )\delta(k^++n\cdot \mathcal
P^\dagger)\) and \(\bar q'_{s,-k^-}=(\bar q'_{s} S_\bn ) \delta(k^- + \bn\cdot
\mathcal P^\dagger)\) and the direction for the soft Wilson lines $S_n$ and
$S_\bn$ are determined by the matching from \SCETa. Just like the local
annihilation operators, we see that the \(O_i^{(1T)}\)'s can not create
transversly polarized vector mesons. The basis for $\Delta S=1$ decays,
$O_{is}^{(1T)}$ switches $\bar d_\bn \to \bar s_\bn$.

Next, we carry out the perturbative matching onto the bases in
Eqs.~(\ref{Q2}) and (\ref{O1T}), and derive the factorization theorem.
The \SCETa weak Hamiltonian with Wilson coefficients \(a_{i}^{hc}\)
for the operators $Q_{id}^{(2)}$ is
\begin{equation} \label{matchQ2}
 H_W =  \frac{4G_F}{\sqrt2}\, (\lambda^{(d)}_u + \lambda^{(d)}_c) \sum_{n,\bn} 
   \int [d\omega_{1}d\omega_{2}d\omega_{3}  d\omega_{4}d\omega_{5}]\, 
    \sum_{i=1-8} a_i^{hc}(\omega_j)\,  Q_{id}^{(2)}(\omega_j) \,.
\end{equation}
Since only the penguin operators $O_{5-8}$ contribute, we pulled out
the common CKM factor with \(\lambda^{(d)}_u = V_{ub}V_{ud}^*\) and
\(\lambda^{(d)}_c = V_{cb}V_{cd}^*\). The analogous result for $\Delta
S=1$ has the same $a_i^{hc}$ coefficients. To match onto the
$a_i^{hc}$ at tree level we first do a spin Fierz on the full theory
$O_{5-8}$ operators to obtain spin structures $P_L\otimes P_R$, and
then compute the graphs in Fig.~\ref{fig:Ann}. Only graphs c) and d)
are nonzero and we find [at $\mu=m_b$]
\begin{align}\label{a1a2}
a_1^{hc}(x, y,\bar y)
  &=\frac{\pi\alpha_s(m_b)}{N_C}\bigg\{\frac{2C_F\,C_5+C_6}{y[x(1-y)-1]}
+\frac{(2C_F-C_A)C_5+C_6}{(1- x )y(1-\bar y)}\bigg\} \,,
\nn\\
a_2^{hc}(x,\bar x, y)
  &=\frac{\pi\alpha_s(m_b)}{N_C}\bigg\{-\frac{(2C_F-C_A)C_5+C_6}
     {\bar x[(1-\bar x)(1-y)-1]}
     -\frac{2C_F\,C_5+C_6}{\bar x y(1 - x)}\bigg\} \,.
\end{align}
The coefficients \(a_{3,4}^{hc}\) are identical to \(a_{1,2}^{hc}\)
respectively with the replacements \(C_{5,6}\to
C_{7,8}\). $a_{5-8}^{hc}$ also begin at \(\mathcal{O}(\alpha_s(m_b))\)
but give \(\alpha_s(\mu_i)\)-suppressed contributions when matched
onto \SCETb, so we do not list their values.  These coefficients are
``polluted'' in that one-loop \(\mathcal O(\alpha_s(m_b)^2)\)
contributions proportional to \(C_{1,2}\) could compete numerically
with the results in Eq.~(\ref{a1a2}). Here $x,\bar x,y,$ and $\bar y$
are defined in Fig.~\ref{fig:Ann}, namely $y=\omega_1/m_b$, $\bar
y=-\omega_4/m_b$, $x=\omega_2/m_b$, $\bar x=-\omega_3/m_b$.  For
$n'=n$ as in \(a_{1,3}\), we have $\bar x=1-x$, but \(\bar
y\equiv\!\!\!\!\!\!\slash \:\, 1-y\) since the momentum is shared
between three \(n-\)collinear partons. Likewise, for $n'=\bn$ as in
$a_{2,4}^{hc}$ we have $\bar y=1-y$ but $\bar
x\equiv\!\!\!\!\!\!\slash\:\, 1-x$.

Having constructed the operators \(Q^{(2)}\) and determined their Wilson
coefficients, it is straightforward to match the time-ordered products
\(T[Q^{(2)} \mathcal L_{\xi q}^{(1)}]\) onto the \SCETb operators \(O_i^{1T}\).
For odd indices $i$ and even indices $i'$ we find that integrating out the hard-collinear quark propagator, shown as the
dashed line inside the gray region in Fig.~\ref{fig:AnnFACT}, gives
\begin{align} \label{matchO1T}
   i \int\!\! d^4x \: T [Q_{id}^{(2)}(\omega_j)](0) \mathcal L_{\xi q}^{(1)}(x) =
  &  \frac{-1}{N_c} 
   \int dk^+ \, 
      O_{id}^{(1T)}(k^+,\omega_{j}) \,,
\nn\\
 i \int\!\! d^4x \: T [Q_{i'd}^{(2)}(\omega_j)](0) \mathcal L_{\xi q}^{(1)}(x) =
  & \frac{-1}{N_c} 
   \int dk^- \, 
       O_{i'd}^{(1T)}(k^-,\omega_{j}) \,.
\end{align}
At ${\cal O}(\alpha_s^2)$ in perturbation theory this matching would
include non-trivial jet functions. For example, in the first line a
$\int d\omega_{1,4}' J(k^+,\omega_{1,4}',\omega_{1,4})$ with
$\omega_{1,4}'$ taking the place of $\omega_{1,4}$ in $O_{id}^{(1T)}$.
However at this order additional time-ordered products and
non-perturbative functions become relevant so we stick to ${\cal
O}(\alpha_s)$ in our analysis. Together
Eqs.\,(\ref{a1a2},\ref{matchO1T}) complete the tree-level matching.
Now take the matrix element of $O_{id}^{(1T)}$ using
\begin{align} \label{lcdistn}
  \langle \pi^+_{n_1}(p) | \bar u_{n,\omega_1}\, \bnslash P_{L,R}\,
  d_{n,\omega_4} | 0 \rangle 
  &=  \frac{\pm i\, f_P }{2}\,  
  \delta_{nn_1}\,
   \delta(\bn\mcdot p \!-\! \omega_1\!+\!
  \omega_4)\, \phi_P(y) \,, \\
 \langle \rho^+_{n_1}(p,\varepsilon) | 
   \bar u_{n,\omega_1}\, \bnslash P_{L,R}\, 
  d_{n,\omega_4} | 0 \rangle &= \frac{i f_V  m_V {\bn\mcdot
      \varepsilon}}{2\,\bn\mcdot p}\, 
   \delta_{nn_1}\,
   \delta(\bn\mcdot p \!-\! \omega_1\!+\! \omega_4)\, \phi_{V_\parallel}(y)
  \,, \nn
\end{align}
and the three-body distributions
\begin{align}
  \langle \pi^+_{n_1}(p) |\, \bar u_{n,\omega_1}\, \bnslash\,
  (i g \slashed{\mathcal B}_\perp)_{n,\omega_5} P_{R}\, d_{n,\omega_4}\, 
   | 0 \rangle 
&  =  \frac{i f_{3P}}{\omega_5}\, 
  \delta_{nn_1}\, 
   \delta(\bn\mcdot p \!-\! \omega_1 \!-\! \omega_5 \!+\! \omega_4)\, 
   \phi_{3P}(y,\bar y) , 
  \\
  \langle \rho^+_{n_1}(p,\varepsilon) |\, \bar u_{n,\omega_1}\, \bnslash\,
  (i g \slashed{\mathcal B}_\perp)_{n,\omega_5} P_{R}\, d_{n,\omega_4}\, 
   | 0 \rangle 
&  =  \frac{i f_{3V} m_V \bar n \cdot \varepsilon}{\omega_5\: \bar n \cdot p}\, 
  \delta_{nn_1}\,
   \delta(\bn\mcdot p \!-\! \omega_1 \!-\! \omega_5 \!+\! \omega_4)\, 
   \phi_{3V}(y,\bar y) \,.\nn
\end{align}
Our convention for the vector meson matrix element has been chosen to simplify
the final result for the amplitude and is related to that
of~\cite{Hardmeier:2003ig} by \(f_{3V} = m_V f_V^T\) and
\(\phi_{3V}=-\mathcal{T}/2\). Permutations in the flavors give the definitions for other meson channels, and we use the phase convention in~\cite{Gronau:1994rj}.  The soft matrix element is
\begin{align}
\langle 0 | \bar q_{s,- n'\cdot k}^{(f)} P_L\, {\slash\!\!\! n' }\,S_{n^\prime}^\dagger b_v | \bar B \rangle 
&= i  \frac{f_B\, m_B\, }{2} \phi_B^+(n'\mcdot k)\,.
\end{align}
\begin{table}[t]
\begin{tabular}{|c|c|c|}
\hline\hline
$B\to M_1 M_2$ & $H_{hc 1} $ & $H_{hc 2} $  
\\ \hline\hline 
  $\pi^0 \pi^-$, $\rho^0\pi^-$ $\pi^0 \rho^-$  &
   $-\frac{1}{\sqrt2}\, a^{hc}_1 -\frac{1}{\sqrt2}\, a^{hc}_3 $
   & $\frac{1}{\sqrt2}\, a^{hc}_2 +\frac{1}{\sqrt2}\, a^{hc}_4 $ \\
  $\pi^- \pi^0$, $\rho^-\pi^0$ $\pi^- \rho^0$
   & $\frac{1}{\sqrt2}\, a^{hc}_1 +\frac{1}{\sqrt2}\, a^{hc}_3 $ &
   $-\frac{1}{\sqrt2}\, a^{hc}_2 -\frac{1}{\sqrt2}\, a^{hc}_4 $ \\
$\pi^+ \pi^-$,  $\pi^+\rho^-$, $\rho^+\pi^-$ 
 &  $- a^{hc}_1  +\frac{1}{2} a^{hc}_3  $ 
  & $a^{hc}_2  -\frac{1}{2} a^{hc}_4 $ 
 \\
$\pi^0 \pi^0,$ $ \rho^0 \pi^0$ 
 & $  a^{hc}_1 -\frac12\, a^{hc}_3   $ 
 &   $- a^{hc}_2 +\frac12\, a^{hc}_4  $ 
  \\
  $\bar K^{0} K^{(*)0}$,  $\bar K^{(*)0} K^{0}$ 
  & $a^{hc}_1 -\frac{1}{2} a^{hc}_3 $ 
  &  $-a^{hc}_2 +\frac{1}{2} a^{hc}_4 $
  \\
 $K^{-} K^{(*)0}$, $K^{(*)-} K^{0}$ 
    & $a^{hc}_1 + a^{hc}_3 $  
    &  $-a^{hc}_2 - a^{hc}_4 $
  \\
\hline
 $\pi^- \bar K^{(*)0}$, $\rho^- \bar{K}^{0}$ 
   & $  a_1^{hc} +a_3^{hc}  $ & $-a_2^{hc} -a_4^{hc}  $
  \\
 $\pi^0 K^{(*)-},\rho^0 K^-$  
   & $-\frac{1}{\sqrt 2}\, a_1^{hc} -\frac{1}{\sqrt 2}\, a_3^{hc}  $ 
   & $\frac{1}{\sqrt 2}\, a^{hc}_2  + \frac{1}{\sqrt 2}\, a^{hc}_4  $ 
   \\
 $\pi^0 \bar K ^{(*)0},\rho^0 \bar K ^{0}$ 
   & $ \frac{1}{\sqrt 2} \, a_1^{hc} -\frac{1}{2\sqrt 2} \, a_3^{hc}  $
   & $-\frac{1}{\sqrt 2} \, a^{hc}_2 +\frac{1}{2\sqrt 2} \, a^{hc}_4  $
   \\
 $\pi^+ K^{(*)-} ,\rho^+ K^{-}$ 
   &  $-a_1^{hc} +\frac{1}{2} a_3^{hc} $ 
   &  $a_2^{hc} -\frac{1}{2} a_4^{hc} $ \\
  \hline\hline
  $B_s \to M_1 M_2$ & $H_{{hc} 1} $ & $H_{{hc} 2} $  
\\ \hline\hline 
$ K^+\pi^-$,  $K^{*+}\pi^- $, $K^+\rho^- $ 
  &  $- a^{hc}_1 +\frac{1}{2} a^{hc}_3   $ 
  &  $a^{hc}_2 -\frac{1}{2} a^{hc}_4  $
 \\
$K^0\pi^0 $, $K^{*0}\pi^0 $, $K^0\rho^0 $
  & $\frac{1}{\sqrt 2}\, a^{hc}_1  -\frac{1}{2\sqrt 2}\, a^{hc}_3   $  &  
  $-\frac{1}{\sqrt 2}\, a^{hc}_2  +\frac{1}{2\sqrt 2}\, a^{hc}_4   $ 
  \\
\hline
 $K^{+}K^{-} $, $ K^{*+}K^{-}$, $K^{+}K^{*-} $
   &  $- a^{hc}_1 +\frac{1}{2} a^{hc}_3 $ & 
  $a^{hc}_2  -\frac{1}{2} a^{hc}_4 $ 
  \\
  $K^{0}\bar K^{0} $,  $K^{*0}\bar K^{0} $ , $ K^{0}\bar K^{*0}$ 
  &  $a^{hc}_1 -\frac{1}{2} a^{hc}_3 $ 
  & $-a^{hc}_2 +\frac{1}{2} a^{hc}_4 $ 
  \\
  \hline\hline
\end{tabular}
\caption{Hard functions for the annihilation amplitude $A_{Tann}^{(1)}$ in
Eq.~(\ref{A1Tann}) for $\bar B^0$, $B^-$, and $\bar B_s$ decays. The
result for $B^- \to \pi^0\pi^-$ is obtained by adding the results
using the entries from the first two rows, and so vanishes in the 
isospin limit.}
\label{tablehca}
\end{table}
Combining these pieces the factorization theorem with tree-level jet
functions is
\begin{align} \label{A1Tann}
A^{(1ann)}_{\rm hard-collin} =\: 
  & \frac{-G_F f_B m_B }{\sqrt{2}\, m_b N_c}\,
  (\lambda^{(d)}_u \!+\! \lambda^{(d)}_c)\!\!
  \int_0^\infty \!\!\!\!\!dk \: \frac{\mathcal \phi_B^+(k)}{k}  \\
&  \times \bigg\{f_{3M_1}f_{M_2} \!\! \int_0^1\!\!\! dx \! \int_0^1\!\!\! dy\! \int_0^{1-y}\!\!\!\!\! d\bar y \ 
  \frac{H_{hc1}^{M_1M_2}(x,y,\bar y)}{1-y-\bar y }\
     \phi_{3M_1}(y ,\bar y ) \phi_{M_2}(x )
  \nn\\
&\quad\  \ +\eta_{M_1}f_{M_1}f_{3M_2}\int_0^1 \!\! dy \! \int_0^1\!\!\! dx \! \int_0^{1-x}\!\!\!\!\!  d\bar x\   
    \frac{H_{hc2}^{M_1M_2}(x,\bar x,y)}{1-x -\bar x}\ \phi_{M_1}(y)
     \phi_{3M_2}(x ,\bar x ) \bigg\} 
     \,, \nn
\end{align}
where \(\eta_{M}=-1\) or \(+1\) for a pseudoscalar or vector meson,
respectively. The hard coefficients $H_{hc1}^{M_1M_2}$ and
$H_{hc2}^{M_1M_2}$ for different $B\to M_1 M_2$ channels are listed in
Table~\ref{tablehca} in terms of coefficients in the \SCETa weak
Hamiltonian.  The amplitude contains the three-body distribution
function as promised. The convolutions in Eq.~(\ref{A1Tann}) are real,
and assuming the standard endpoint behavior for the distribution
functions they converge without the rapidity factorization
of~\cite{Manohar:2006nz}.

We conclude by comparing our result parametrically and numerically to
$A^{(1ann)}_{\rm hard}$ and $A^{(2ann)}_{\rm hard}$ as defined in
Ref.~\cite{Arnesen:2006vb}. For this comparison it is useful to define
moment parameters
\begin{align} \label{betahc}
  \beta_{\hc 1}^{M_1 M_2} , \, \beta_{\hc 3}^{M_1 M_2}  &= \int \!\! dx dy d\bar y\:
   \frac{a_{1,3}^{\hc}(x,y,\bar y)}{1-y-\bar y}\, \phi_{3M_1}(y,\bar y)
   \phi_{M_2}(x)    \,,
  \\
\beta_{\hc 2}^{M_1 M_2} ,\, \beta_{\hc 4}^{M_1 M_2}  &= \int \!\! dy dx d\bar x \:
   \frac{a_{2,4}^{\hc}(x,\bar x,y)}{1-x-\bar x}\, \phi_{M_1}(y) \phi_{3M_2}(x,\bar x) \,,
   \qquad
   \beta_B = \frac{1}{3} \int \frac{dk}{k}  \phi_B^+(k) \nn \,,
\end{align}
where $\beta_B = \lambda_B^{-1}/3$ has mass dimension $-1$. First we
compare the leading-power annihilation amplitudes in \(\bar B\to\pi^+
K^-\). Dropping terms proportional to the tiny Wilson coefficients
\(C_{7-8}\), we have
\begin{align}\label{R1}
 R_1({ \pi^+ K^-})\equiv\frac{ A_{\rm hard-col}^{(1ann)}(\pi^+ K^-) }
 { A_{\rm hard}^{(1ann)}(\pi^+ K^-) }&=  
   \frac {  \frac{G_F f_B m_B}{\sqrt{2} m_b N_c} (\lambda^{(s)}_u \!+\!\lambda^{(s)}_c)\: 3 \beta_B
       \big[ {-} f_{3\pi} f_K \beta_{hc1}^{K \pi} +f_{3K} f_\pi \beta_{hc2}^{K \pi} \big] }
      { \frac{G_F f_B  }{\sqrt{2} } (\lambda^{(s)}_u\!+\!\lambda^{(s)}_c)
         f_\pi f_K \big[  {-}\beta_{4u}^{K \pi} \big] }\,.
\end{align}
Parametrically, the moments in \(R_1\) have
\(\beta_{4u}\sim\beta_{hci}\sim\mathcal O(\alpha_s(m_b))\), and the
power counting of the prefactor is \(f_{3K}\beta_B/f_K\sim 1\). Also
there is no suppression from the hierarchy in the \(C_i\)'s since
\(\beta_{4u}\) involves \(C_3\), and \( C_3\approx C_5\approx C_6\).
Thus, we have shown that for consistency in the \(\alpha_s\) and
\(1/m_b\) expansion, the contributions \(A^{(1ann)}_{hard-col}\) need
to be included with the local contributions \(A^{(1ann)}_{hard}\) in
the leading annihilation amplitude. Similarly we can compare the new
hard-collinear annihilation amplitude to the chirally enhanced
annihilation contribution in $\bar B^0 \to
\pi^+K^-$. Isolating the terms proportional to the large coefficients
$C_5$ and $C_6$ we have
\begin{align}\label{R2}
 R_2({\pi^+ K^-})\equiv\frac{ A_{\rm hard-col}^{(1ann)}(\pi^+K^-) }{ A_{\rm hard}^{(2ann)}(\pi^+K^-) }&=  
   \frac {  \frac{G_F f_B m_B}{\sqrt{2} m_b N_c} (\lambda^{(s)}_u \!+\!\lambda^{(s)}_c)\: 3 \beta_B
       \big[ {-} f_{3\pi} f_K \beta_{hc1}^{\pi K}- f_{\pi} f_{3K} \beta_{hc2}^{\pi K} \big] }
      { \frac{G_F f_B   }{\sqrt{2} m_b } (\lambda^{(s)}_u\!+\! \lambda^{(s)}_c)
         f_\pi f_K
  \big[ {-} \mu_\pi \beta_{\chi 1}^{\pi K}+\mu_K \beta_{\chi 2}^{\pi K} \big] }
  .
\end{align}
Parametrically \(\beta_{\chi }\sim\beta_{hc }\sim \alpha_s(m_b)\), and
\(R(\pi^+K^-)\sim{m_B\beta_B f_{3\pi}}/({f_\pi\mu_\pi})\sim
m_b/\mu_\pi\sim m_b/\Lambda\) as expected. 

We conclude with a brief numerical analysis of the ratios \(R_1\) and
\(R_2\). The $C_i$'s are quoted below Eq.~(\ref{fullops}), and we use
\(\alpha_s(m_b) = 0.22,\, f_K = 0.16\,{\rm GeV},\,\) and 
\( f_\pi =0.13\,{\rm GeV}\). $f_B=0.22\,{\rm GeV}$ is taken from a 
recent lattice determination~\cite{Gray:2005ad}, the three-body decay
constants \(f_{3K} \simeq 4.5 \times 10^{-3}\, {\rm GeV}^2\) and \(
f_{3\pi} \simeq 4.5 \times 10^{-3}\, {\rm GeV}^2\) come from QCD sum
rules~\cite{Ball:2006wn} and $\beta_B \simeq 1/(.4\,{\rm GeV})$ was
determined in a fit to nonleptonic data~\cite{Bauer:2005kd}.  To model
the nonperturbative meson distributions we truncate the conformal
partial wave expansions~\cite{Zhitnitsky:1985dd} as
\begin{align}\label{phimodels}
  \phi^M(x) &= 6x(1-x) \big[1+ a_1^M (6x-3) + 6 a_2^M ( 1- 5 x+
  5x^2)\big] \,,\nn\\ \phi_{3M}(x,\bar x) &=360 x \bar x (1-x-\bar
  x)^2 \big[1+\frac{w_{3M}}{2}\{7(1-x-\bar x)-3\}\big] \,.
\end{align}
Eq.~(\ref{betahc}) has convergent convolution integrals for these
distribution functions.  To estimate the moments \(\beta\)
and the ratios \(R\) we vary the coefficients in Eq.~(\ref{phimodels})
in a conservative range inferred from recent lattice
results~\cite{Braun:2006dg} for the $a_i^M$'s and QCD sum
rules~\cite{Ball:2006wn} for the \(w_{3M}\)'s. Specifically we take
\(a_1^\pi=0,\, a_1^K=0.05\pm 0.02,\, a_2^{\pi,K}=0.2\pm 0.2,\) and 
\(w_{3\pi,K}=-1\pm 1\). A Gaussian scan of the model parameters gives
\begin{align}\label{numbers}
\beta ^{\bar K K}_{hc 1}&=-1.4 \pm 0.4, & \beta^{\bar K K}_{hc2}&=0.3\pm 0.1, & 
R_1(\pi^+ K^-)&= 0.3 \text{ to }\, 1.2 \,,
\nn\\
\beta ^{\pi K}_{hc 1}&=-1.4 \pm 0.5, &  \beta^{\pi K}_{hc2}&=0.1\pm 0.1, 
& R_2({\pi^+ K^-})&=-0.1 \text{ to } 0.1 \:.
\end{align}
The denominators of Eq.~(\ref{R1})~and~(\ref{R2}) can vanish, giving
large departures from Gaussian statistics. So for \(R_1\) and \(R_2\)
we quote the range that contains an equivalent number of points as one
standard deviation for a Gaussian distribution. Eq.~(\ref{numbers})
demonstrates that numerically the three-parton contributions to
\(A^{(1ann)}\) could be of the same size or larger than the local
piece
\(A^{(1ann)}_{hard}\). Numerically, ${m_B\beta_B
f_{3\pi}}/({f_\pi\mu_\pi})\sim 0.2$ causing some suppression in
$R_2(\pi^+K^-)$. It would be interesting to examine the size of these
three-parton contributions in the $k_T$-approach of Ref.~\cite{Keum:2000ph}.

In this paper we computed the final missing term of the leading order
annihilation amplitude in \(B\to M_1 M_2\) decays. These terms involve
a three-parton distribution and need to be included for a complete
analysis of annihilation.

\acknowledgments

We thank Z.Ligeti for discussion. This work was supported in part by
the Office of Nuclear Physics of the U.S.\ Department of Energy under
the cooperative research agreement DOE-FC02-94ER40818 (C.A. and I.S.),
and DOE-ER-40682-143 and DEAC02-6CH03000 (I.R.).  I.S.~was also
supported in part by the DOE OJI program and by the Sloan Foundation.

\end{document}